\documentclass[10pt,twocolumn,letterpaper]{article}

\usepackage{iccv}
\usepackage{times}
\usepackage{epsfig}
\usepackage{graphicx}
\usepackage{physics}
\usepackage{cancel}

 \usepackage{amssymb}

{\begin{list}               
    {$\bullet$ \hfill}{
        \setlength{\leftmargin}{\parindent}
        \setlength{\parsep}{0.04\baselineskip}
        \setlength{\itemsep}{0.5\parsep}
        \setlength{\labelwidth}{\leftmargin}
        \setlength{\labelsep}{0em}}
    }
{\end{list}}

\providecommand{\eref}[1]{\eqref{#1}}  
\providecommand{\cref}[1]{Chapter~\ref{#1}}

\providecommand{\fref}[1]{Figure~\ref{#1}}

\providecommand{\R}{\ensuremath{\mathbb{R}}}

\renewcommand{\vec}[1]{\ensuremath{\boldsymbol{#1}}}
\providecommand{\mat}[1]{\ensuremath{\boldsymbol{#1}}}

\providecommand{\mH}{\mat{H}}

\providecommand{\mR}{\mat{R}}

\providecommand{\vh}{\vec{h}}

\providecommand{\vx}{\vec{x}}
\providecommand{\vy}{\vec{y}}

\providecommand{\valpha}{\vec{\alpha}}
\providecommand{\vbeta}{\vec{\beta}}

\providecommand{\vrho}{\vec{\rho}}

\providecommand{\vvarphi}{\vec{\varphi}}

\usepackage[pagebackref=true,breaklinks=true,letterpaper=true,colorlinks,bookmarks=false]{hyperref}

\iccvfinalcopy 

\def\iccvPaperID{4332}

\ificcvfinal\fi

\begin{document}

\title{Accelerating Atmospheric Turbulence Simulation via Learned \\ Phase-to-Space Transform}

\author{Zhiyuan Mao, Nicholas Chimitt, Stanley H. Chan\\
School of Electrical and Computer Engineering, Purdue University, West Lafayette, Indiana USA\\
{\tt\small \{mao114,nchimitt,stanchan@purdue.edu\}}
}

\maketitle
\ificcvfinal\thispagestyle{empty}\fi

\begin{abstract}
Fast and accurate simulation of imaging through atmospheric turbulence is essential for developing turbulence mitigation algorithms. Recognizing the limitations of previous approaches, we introduce a new concept known as the phase-to-space (P2S) transform to significantly speed up the simulation. P2S is built upon three ideas: (1) reformulating the spatially varying convolution as a set of invariant convolutions with basis functions, (2) learning the basis function via the known turbulence statistics models, (3) implementing the P2S transform via a light-weight network that directly converts the phase representation to spatial representation. The new simulator offers $300\times$ -- $1000\times$ speed up compared to the mainstream split-step simulators while preserving the essential turbulence statistics. 
\end{abstract}

\section{Introduction}
Despite several decades of research, imaging through atmospheric turbulence remains an open problem in optics and image processing. The challenge is not only in reconstructing images from a stack of distorted frames but also in a less known image formation model that can be used to formulate and evaluate image reconstruction algorithms such as deep neural networks. 
Simulating images distorted by atmospheric turbulence has received considerable attention in the optics community \cite{SchmidtTurbBook, RoggemannSimulator, Hardie2017, Potvin_Forand_Dion_2011}, but using these simulators to develop deep learning image reconstruction algorithms remains a challenge as there is no physically justifiable approach to synthesize large-scale datasets at a low computational cost for training and testing.

\begin{figure}[th]
\centering
\begin{tabular}{cc}
\multicolumn{2}{c}{\hspace{-2ex}\includegraphics[width=\linewidth]{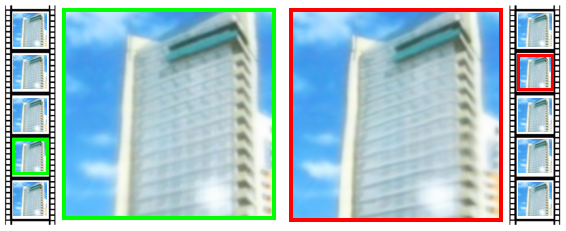}}\\
24.36 sec / frame (GPU)  & 0.026 sec / frame (GPU)\\
(a) Hardie et al. \cite{Hardie2017} & (b) Ours
\end{tabular}
\caption{This paper presents a new turbulence simulator that is substantially ($1000\times$) faster than the prior art, while preserving the essential turbulence statistics.}
\label{fig: pos_paper}
\end{figure}

\begin{figure}[th]
	\centering
	\begin{tabular}{ccc}
		\hspace{-1ex}\includegraphics[width=0.32\linewidth]{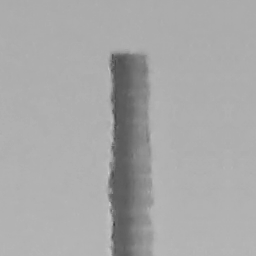}& \hspace{-2ex}\includegraphics[width=0.32\linewidth]{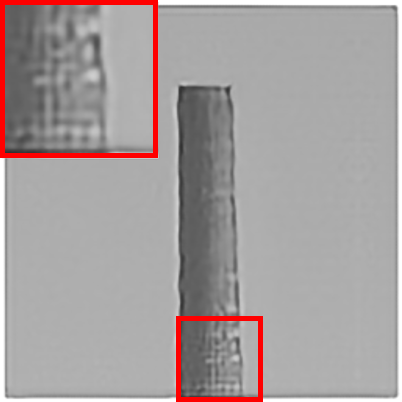}&
		\hspace{-2ex}\includegraphics[width=0.32\linewidth]{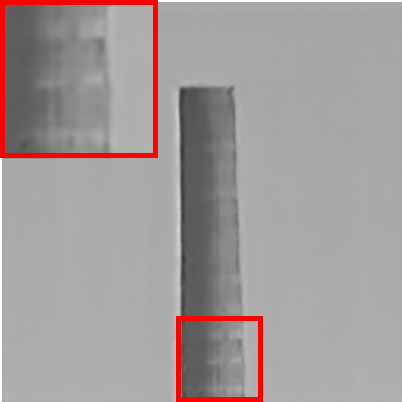}\\
		(a) Input (real) & \hspace{-2ex}(b) \cite{Lau2021_sim}+U-Net & \hspace{-2ex}(c) Ours+U-Net
	\end{tabular}
	\caption{Using our simulator to synthesize training set for training an image reconstruction network (U-Net \cite{Unet}) offers a considerable amount of improvement in image quality. The network is identical for both (b) and (c); only the simulator used to synthesize the training data is different.}
	\label{fig: recon_wow}
	\vspace{-2ex}
\end{figure}

Recognizing the demand for a fast, accurate, and open-source simulator, we present a new method to generate a dense-grid image distorted by turbulence with theoretically verifiable statistics. The simulator consists of mostly optics/signal processing steps and a lightweight shallow neural network to perform a new concept called the \textbf{Phase-to-Space (P2S)} transform. By parallelizing the computation across the pixels, the simulator offers a $1000\times$ speed-up compared to the mainstream approach as shown in \fref{fig: pos_paper}. When using the new simulator to synthesize training data to train a deep neural network image reconstruction model, the resulting network outperforms the same architecture trained with data synthesized by a less sophisticated simulator, as illustrated in \fref{fig: recon_wow}.

An overview of the proposed simulator is illustrated in \fref{fig:overall idea}. Our proposed approach is based on linking the following two ideas:
\begin{figure}[ht]
    \centering
    \includegraphics[width=\linewidth]{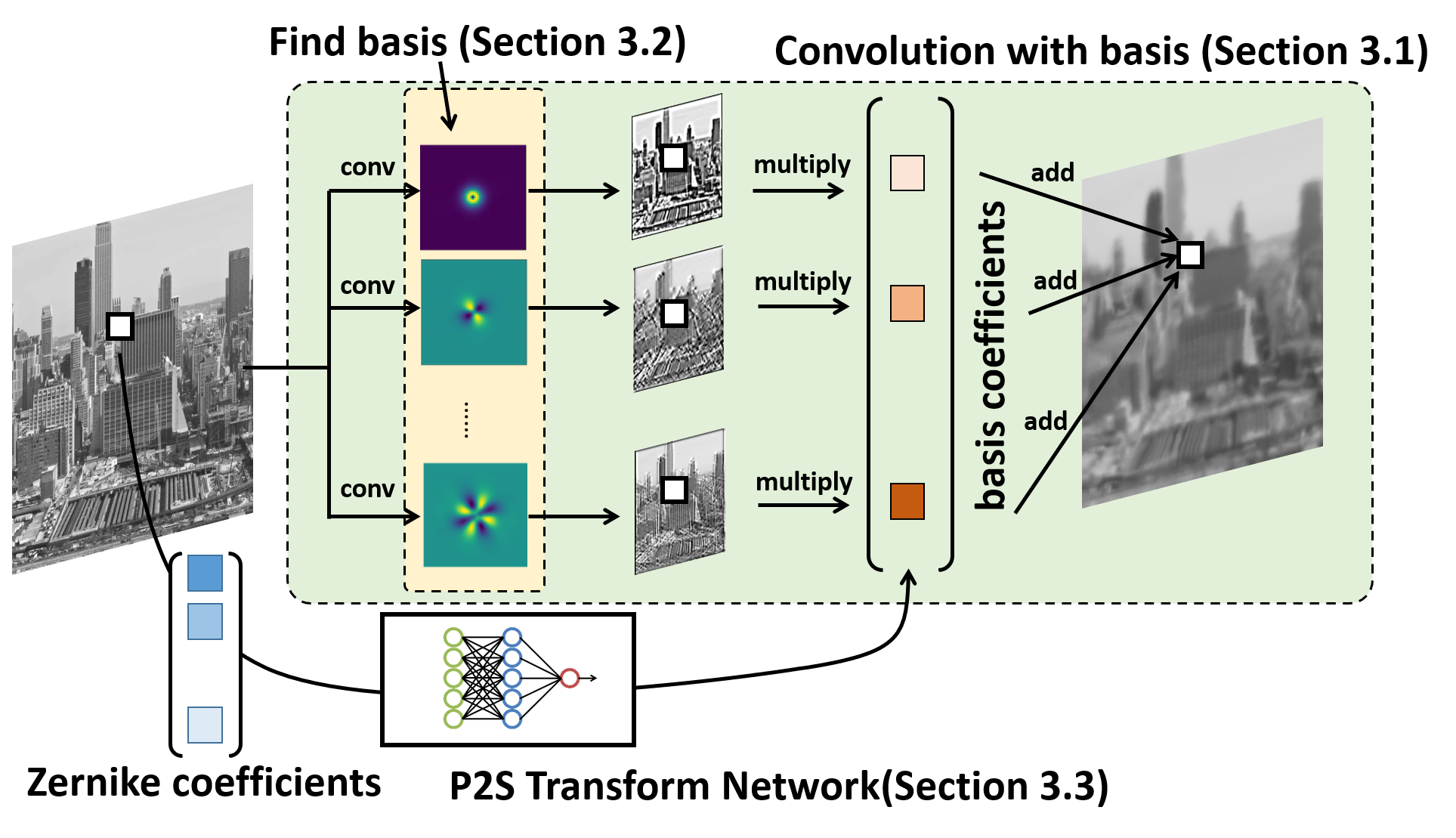}
    \caption{This paper introduces three ideas to significantly speed up the simulation. The three ideas are: (Section 3.1) Approximating the spatially varying convolution by invariant convolutions, (Section 3.2) learning the basis representation via known turbulence statistics, (Section 3.3) implementing the Phase-to-Space transform network.}
    \label{fig:overall idea}
\end{figure}

\begin{itemize}
\setlength\itemsep{0ex}
\item \textbf{Convolution via basis functions} (Section 3.1). While conventional approaches model the turbulence distortion as a spatially varying convolution, we reformulate the problem by modeling the distortion as a sum of spatially invariant convolutions. The idea is to utilize a basis representation of the point spread functions (PSFs). This concept is similar to the prior work of \cite{spatvarconv}, but in a different context.

\item \textbf{Learning the basis functions} (Section 3.2). To enable the previous idea, we need to have the basis functions. This is done by utilizing \cite{Chimitt2020} to draw Zernike samples for all \emph{high-order} aberrations. Then, principal component analysis is used to construct the basis functions as proposed by Mao et al. \cite{mao_tci}. This is also reminiscent to the dictionary approach proposed by Hunt et al. \cite{Hunt_Iler_SparseRepresntation}.
\end{itemize}

The missing piece between these two ideas is the relationship between the basis coefficients in the phase and spatial domains. This is an open problem, and there is no known analytic solution. We circumvent this difficulty by introducing a new concept known as the \textbf{Phase-to-Space transform} (Section 3.3). To do so, we construct a lightweight shallow neural network to transform from the phase domain to the spatial domain. Integrating this network into the two aforementioned ideas, our overall simulator adheres to the physics while offering significant speed up and additional reconstruction utility.

\section{Background}
In this section we provide a brief summary of the turbulence physics and prior work in turbulence simulation. The theory of imaging through atmospheric turbulence can be traced back to the work of Kolmogorov \cite{Kolmogorov1941} and Tatarski \cite{Tatarski1967}, followed by a series of major breakthroughs by Fried \cite{Fried65,Fried66optical,Fried78} and Noll \cite{Noll_1976}. Readers are encouraged to check out \cite{roggemann1996imaging, Goodman_StatisticalOptics} for an introduction.

\begin{figure}[ht]
\centering
\includegraphics[width=\linewidth]{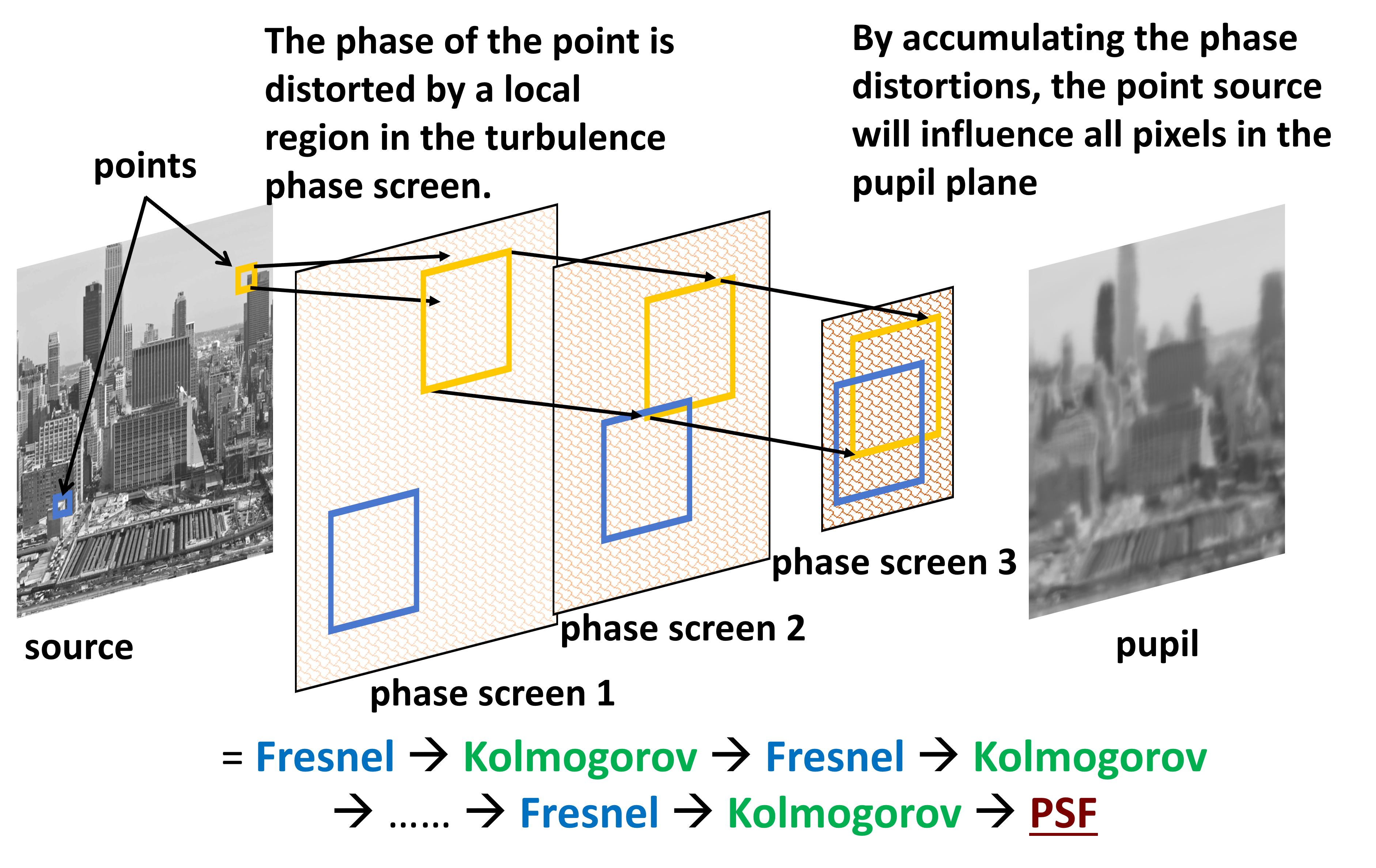}
\caption{Split-step propagation \cite{Hardie2017} models the turbulence as a discrete set of phase screens where the wavefront distortion is caused by cropping regions of the phase screen at \emph{every} pixel location. The key operations are Fresnel propagation and Kolmogorov phase imparting. The end result of a sequence of these operations is a PSF for one pixel. The overlaps of the phase screens create the spatial correlations. See \cite{Hardie2017} for detailed description.}
\label{fig: phase screen}
\end{figure}

\subsection{Split-step simulation}

The image formation process through turbulence is best described in the \emph{phase} domain. In free space, an emitted wave propagates spherically outward and, if at a sufficiently long distance, arrives upon the aperture approximately flat. If the medium contains random fluctuations, the phase of the wavefront will be distorted along the path of propagation. We can imagine the wave leading and lagging in phase in reference to its unperturbed counterpart as a result of spatially varying indices of refraction.

\begin{figure}[ht]
    \centering
    \includegraphics[width=0.96\linewidth]{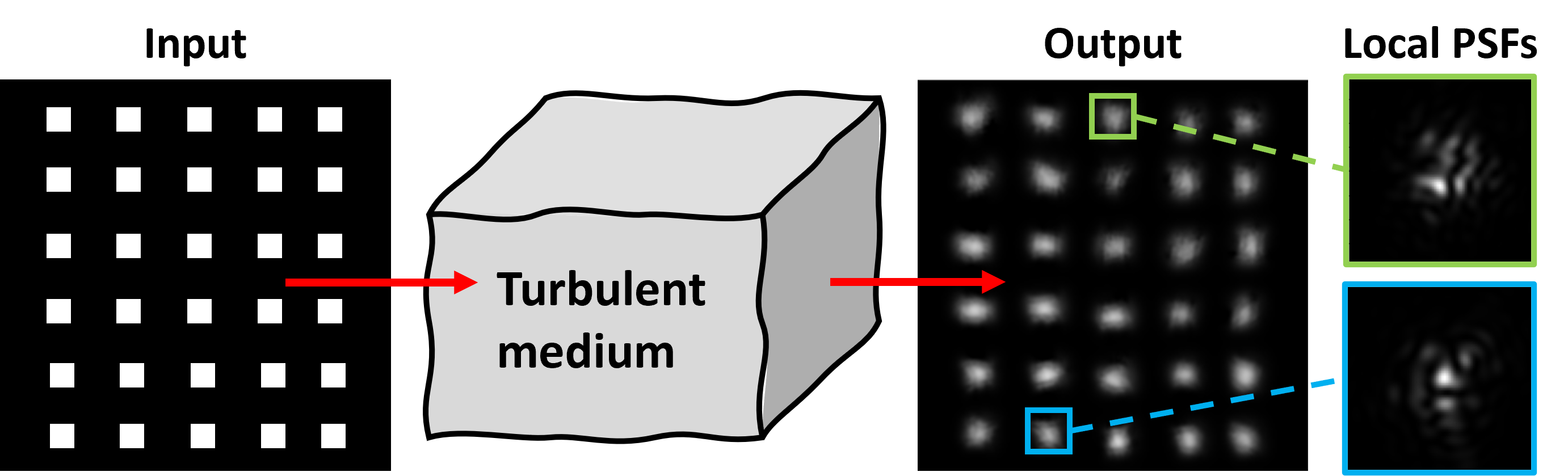}
    \caption{If we image a grid of point sources through turbulence, we will observe a set of spatially varying point spread functions (PSFs). The shape and orientation of the PSFs are determined by the phase structure of the turbulence.}
    \label{fig: image_formation_figure}
\end{figure}

The most widely used simulation approach to the above process is the split-step propagation \cite{SchmidtTurbBook, RoggemannSimulator,Hardie2017}. The idea is to discretize the wave propagation path as illustrated in \fref{fig: phase screen}. Split-step simulation propagates every point in the object plane through a discrete set of \emph{phase screens}, alternating between free space propagation, given by Fresnel diffraction, and phase imparting. The statistical behavior of the phase screens is defined through its power spectral density (PSD) \cite{Hardie2017, SchmidtTurbBook}, many of which are related to the Kolmogorov PSD. This sequence of operations is best described by the equation
\begin{equation*}
\text{Fresnel} \rightarrow \text{Kolmogorov} \rightarrow \ldots \rightarrow \text{Fresnel} \rightarrow \text{Kolmogorov}.
\end{equation*}
After passing through a turbulent medium, the point spread functions (PSFs) will be spatially varying as illustrated in \fref{fig: image_formation_figure}.

The benefit of split-step is two fold: (1) it is interpretable, as it mirrors the physical process, (2) spatial corrrelations are obtained with minimal effort, as neighboring point sources share overlapping cropped phase screens. The drawback of split-step propagation is its computational requirements: each Fresnel propagation requires a pair of Fourier transforms. This is repeated for every point and every step along the path. Moreover, performing the spatially varying convolution adds another layer of computational cost \cite{Hardie2017,SchmidtTurbBook}.

\subsection{Phase-over-aperture simulation}
\begin{figure}[th]
\centering
\includegraphics[width=\linewidth]{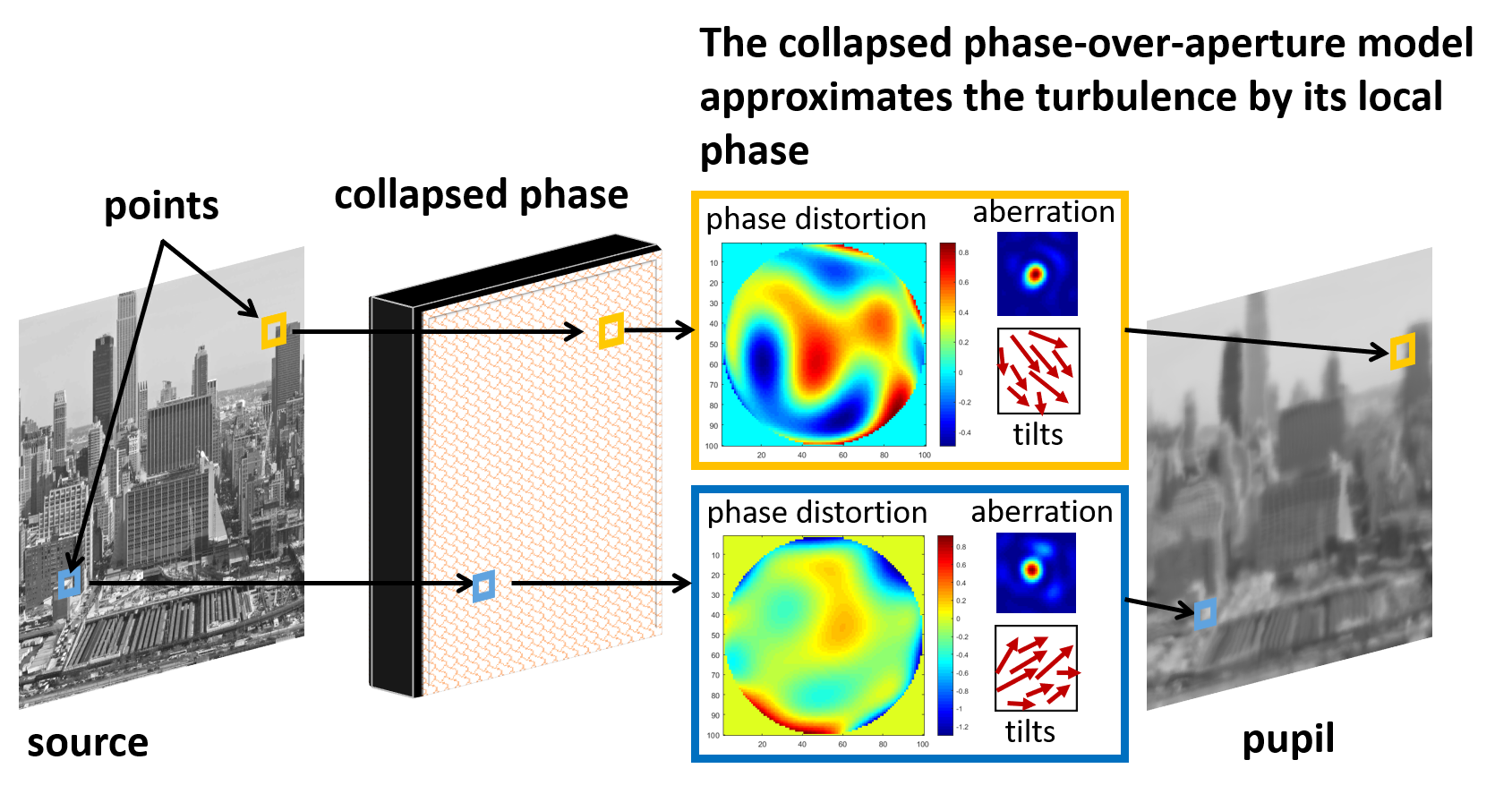}
\caption{The collapsed phase-over-aperture model \cite{Chimitt2020} replaces the global phase screens and Fresnel diffraction by local phase screens per pixel. This translates the wave propagation to a spatially varying convolution with PSFs that are characterized by the tilts and aberrations \emph{per-pixel}. While the phase cropping and propagation of the split-step method is eliminated, Fourier transforms at every pixel location are stilled needed.}
\label{fig: per_pixel_phase}
\end{figure}

Our proposed simulator is inspired by the work of Chimitt and Chan \cite{Chimitt2020}. The idea is to collapse the split-step propagation into the resultant phase across the aperture. Compared to split-step which uses \emph{global} phase screens, the collapsed model generates the \emph{local} phase realization directly, which we illustrate in \fref{fig: per_pixel_phase}.

In the collapsed model, the local per-pixel phase is generated using Noll's idea \cite{Noll_1976} that the phase $\phi(\vrho)$ (defined over the aperture of diameter $D$ with $\vrho$ being the coordinate) can be represented via the Zernike basis functions
\begin{equation}
\phi(\vrho) = \sum_{j=1}^K \alpha_j Z_j(\vrho),
\end{equation}
where $Z_j(\vrho)$ is the Zernike basis and $\alpha_j$ are the Zernike coefficients $[ \alpha_1, \alpha_2, ... \alpha_K ] = \vec{\alpha} \sim \mathcal{N}(\vec{0}, R_Z)$, with \cite{Noll_1976} providing the expression for $R_Z$. The resultant incoherent PSF is formed via
\begin{equation}
    \vh = 
    \left|\mathcal{F} \left\{W(\vrho) e^{-j \phi(\vrho)} \right\}\right|^2,
    \label{eq: PSF_formation}
\end{equation}
omitting a few constants for brevity, with $W(\vrho)$ as the pupil function of the aperture. 

The Zernike representation offers a natural grouping of terms as suggested in \fref{fig: per_pixel_phase}: tilt and higher order aberrations. The terms $\alpha_2$ and $\alpha_3$ correspond to the horizontal and vertical tilt of the plane of best fit to the phase distortion $\phi$. The terms $\alpha_4, \alpha_5, \dots$ correspond to the higher order aberrations and account for the complicated distortions the phase of the wave exhibits. Computationally, these two groups can be separated by generating the high order aberrations, applying the resultant PSFs to the image, then locally shifting the image according to its tilt statistics.

A technical challenge of the collapsed model is ensuring the Zernike coefficients are also \emph{spatially correlated}. In \cite{Chimitt2020}, this correlation is enabled through the invention of a multi-aperture approximation in which the correlations could be described analytically by leveraging several classic works \cite{Basu_2015, chanan92, Takato1995}. With the correlation matrix defined, the spatially correlated tilts can be generated. For the higher-order terms, it was suggested in \cite{Chimitt2020} that one can define a grid of PSFs and \emph{spatially interpolate} between them.

\subsection{Limitations of phase-over-aperture}
As reported in \cite{Chimitt2020}, the collapsed model is significantly faster than the standard split-step propagation. However, by evaluating the simulator, it is evident that there are several limiting factors:
\begin{itemize}
\setlength\itemsep{0ex}
\item The collapsed model exclusively draws Zernike coefficients to create the distortion. However, even with all Zernike coefficients available, one still needs to convert them to PSFs through \eqref{eq: PSF_formation} at \emph{every pixel}. This is the biggest bottleneck.
\item It was suggested that in order to reduce the number of Fourier transforms, one can construct the PSFs for a grid of points, then interpolate between them \emph{spatially}. However, mathematically this is incorrect, as the superposition in the spatial domain is not the same as superposition in the phase domain. 
\item Even if we can resolve the above two problems, to finally simulate a distorted image, we still need to perform the spatially varying convolution. This involves storing the PSFs, and executing the convolution, both of which are resource demanding. 
\end{itemize}

\subsection{Other simulators}
\textbf{Ray Tracing}. An alternative to the split-step simulation is ray tracing \cite{Potvin_Forand_Dion_2011,Lachinova2017}, which requires tracing each point sources through the propagation medium. There are also ray tracing techniques developed in computer graphics \cite{Schwarzman2017ICCP}. However, the lack of quantitative evaluation based on turbulence statistics makes it difficult to assess these methods. 

\textbf{Warp-and-blur}. For faster simulations, one can compromise the accuracy by simulating only the pixel-shifts, commonly referred to as \emph{tilts}, and assuming a spatially-invariant blur \cite{Repasi_Weiss, Leonard_Howe_Oxford}. These simulations and models are widely used in the image processing literature \cite{Milanfar2013, Lau2017, Anantrasirichai2013,Lou2013}, where the goal was to provide quick evaluations of the reconstruction algorithms. However, these methods fail to match the known statistical behavior of the distortions.

\section{Method}
The paper includes two key building blocks: (1) re-formulating the spatially varying convolution via a set of spatially invariant convolutions, (2) constructing the invariant convolutions by learning the basis functions. The major invention here is the linkage between the two for which we introduce the P2S transform to convert the Zernike coefficients to the PSF coefficients.

\subsection{Idea 1: Convolution via basis functions}
The turbulent distortions can be modeled as a spatially varying convolution at each pixel. Denoting $\vx \in \R^N$ as the source image, and $\vy \in \R^N$ as the pupil image, the spatially varying convolution says that $\vy$ is formed by
\begin{equation}
\vy  = 
\begin{bmatrix}
y_1\\
\vdots\\
y_N
\end{bmatrix} = \mH\vx = \begin{bmatrix}
\vh_1^T \vx\\
\vdots\\
\vh_N^T \vx
\end{bmatrix},
\label{eq: convolution 1}
\end{equation}
where $\{\vh_n \,|\, n = 1,\ldots,N\}$ are the $N$ spatially varying PSFs stored as rows of the linear operator $\mH \in \R^{N \times N}$.

The first key idea of the paper is to write 
$\vh_n$ as 
\begin{equation}
\vh_n = \sum\limits_{m=1}^M \beta_{m,n} \vvarphi_m,
\end{equation}
for some basis functions $\vvarphi_m$ (to be discussed) of the PSFs, and coefficients $\beta_{m,n}$ of the $m$\textsuperscript{th} basis at the $n$\textsuperscript{th} pixel. Then, each pixel $y_n$ in \eref{eq: convolution 1} can be written as
\begin{equation}
y_n
= \sum\limits_{m=1}^M \beta_{m,n} \; \vvarphi_m^T\vx, \;\; n = 1,\ldots,N.
\label{eq: convolution 2}
\end{equation}
Since convolution is linear, this turns the $N$ spatially \emph{varying} convolutions $\{\vh_n^T\vx\}_{n=1}^N$ in \eref{eq: convolution 1} into $M$ spatially \emph{invariant} convolutions $\{\vvarphi_m^T\vx\}_{m=1}^M$ in \eref{eq: convolution 2}. If $M \ll N$, the computational cost of \eref{eq: convolution 2} can be much lower. 

To enable the convolution using the basis functions, there are two quantities we need to learn from the data. These are the basis functions $\vvarphi_m$ and the coefficients $\beta_{m,n}$. If we are able to find both, the image can be formed by a simple multiply-add between the basis convolved images $\vvarphi_m^T\vx$ and the representation coefficients $\beta_{m,n}$, as illustrated in \fref{fig:overall idea}.

\subsection{Idea 2: Learning the basis functions}
\begin{figure}[ht]
    \centering
    \includegraphics[width=\linewidth]{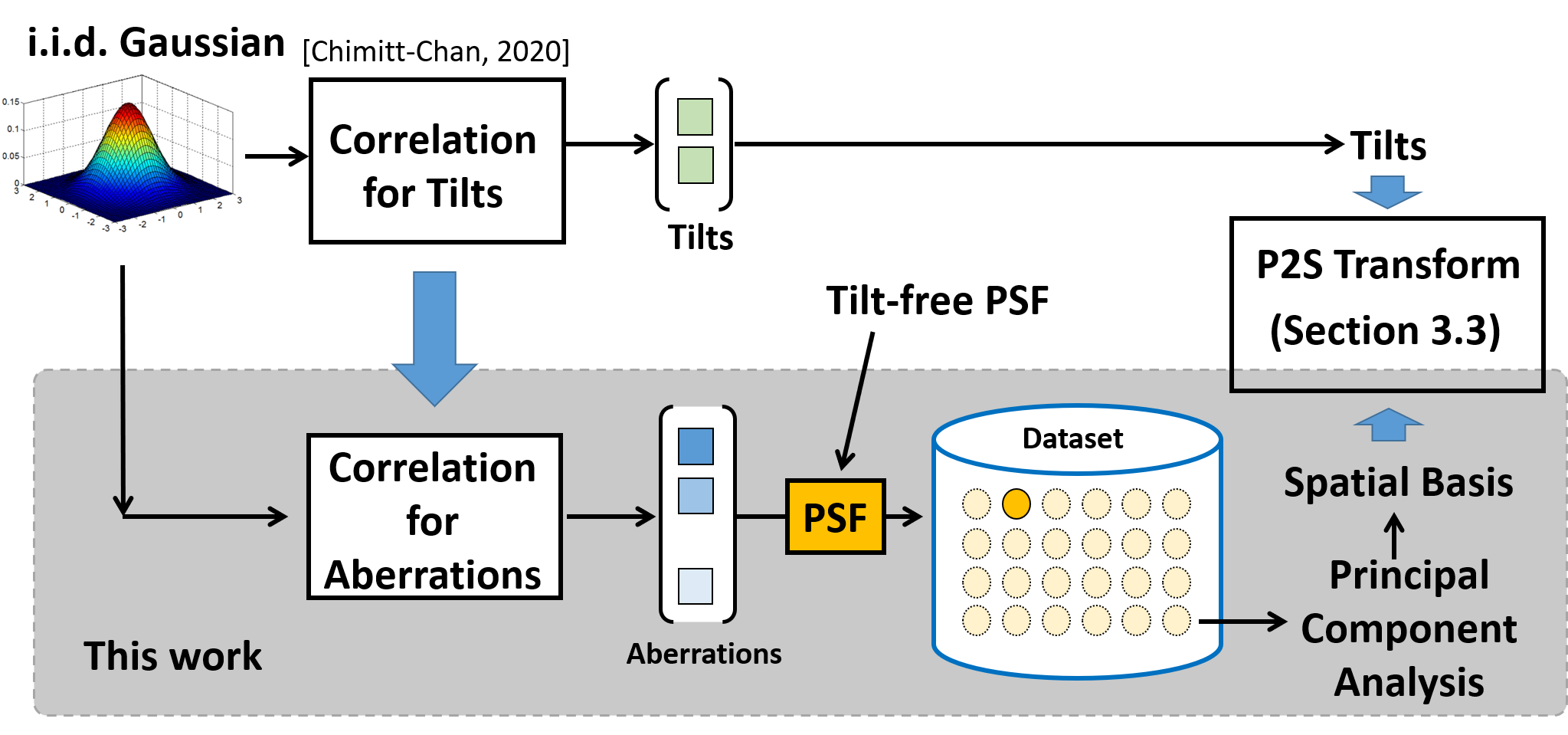}
    \caption{The basis representation is generated in two different ways. For the tilts, we follow the work of \cite{Chimitt2020} to draw spatially correlated tilts by multiplying an i.i.d. Gaussian vector with the tilt correlation matrix. For the high-order aberration terms, we consider the multi-aperture concept of \cite{Chimitt2020} and the analytic solution derived in \cite{Takato1995}. Principal component analysis is conducted to extract the spatial basis functions.}
    \label{fig:basis}
\end{figure}

To generate the basis functions $\vvarphi_m$, we consider the process described in \cite{Chimitt2020} of forming a zero-mean Gaussian vector with a covariance matrix $\mR_Z$ from \cite{Noll_1976}. The strength of correlation is dictated by the optical parameters as well as the relationship $D/r_0$, where $D$ is the aperture diameter and $r_0$ is the Fried parameter \cite{Fried66optical}. \fref{fig:basis} (the upper half) illustrates the generation of the tilts; removing these does not change the shape of the PSF, but instead centers it. We then seek a basis representation of the resulting centered PSFs, which we show in the lower half of \fref{fig:basis}.

To generate the basis functions $\{\vvarphi_m\}_{m=1}^M$, we use the above procedure to construct a dataset containing 50,000 PSFs from weak to strong turbulence levels. (See supplementary material for details.) Given the dataset, we perform a principal component analysis. For the numerical experiments reported in this paper, a total of $M = 100$ basis functions were used. The basis functions are then combined with the tilts, and are sent to the phase-to-space (P2S) transform to determine the basis coefficients $\{\beta_{m,n}\}$.

\subsection{Idea 3: Phase-to-Space (P2S) transform}
The third idea, and the most important one, is the phase-to-space transform. The goal is to define a \emph{nonlinear mapping} that converts the per-pixel Zernike coefficients $\valpha = [\alpha_{1},\ldots,\alpha_K]$ to their associated PSF basis coefficients $\vbeta = [\beta_1,\ldots,\beta_M]$, where we've dropped the pixel index subscript $n$ for notational clarity.

At the first glance, since the basis functions $\{\vvarphi_m\}_{m=1}^M$ are already found, a straightforward approach is to project the PSF $\vh$ (which is defined at each pixel location) onto $\{\vvarphi_m\}_{m=1}^M$. However, doing so will defeat the purpose of skipping the retrieval of $\vh$ from the Zernike coefficients as this is the computational bottleneck. One may also consider \emph{analytically} describing the PSF in terms of $\vvarphi_m$ and the Zernike coefficients,
\begin{equation}
    \vh = 
    \left|\mathcal{F} \left\{W(\vrho) e^{-j \phi(\vrho)} \right\}\right|^2
    \overset{?}{=} \sum_{m=1}^M \beta_m \vvarphi_m.
    \label{eq: PSF_formation_question}
\end{equation}
However, doing so (i.e., establishing the equality in \eref{eq: PSF_formation_question} by writing an equation for $\beta_m$) is an open problem. Even if we focus on a special case with just a single Zernike coefficient, the calculation of the basis functions will involve non-trivial integration over the circular aperture \cite{Goodman_FourierOptics}. 

To bypass the complication arising from \eref{eq: PSF_formation_question}, we introduce a computational technique. The idea is to build a \emph{shallow} neural network to perform the conversion from $\valpha \in \R^K$ to $\vbeta \in \R^M$. We refer to the process as the phase-to-space transform and the network as the P2S network, as the input-output relationship is from the phase domain to the spatial (PSF) domain.

\begin{figure}[ht]
    \centering
    \includegraphics[width = \linewidth]{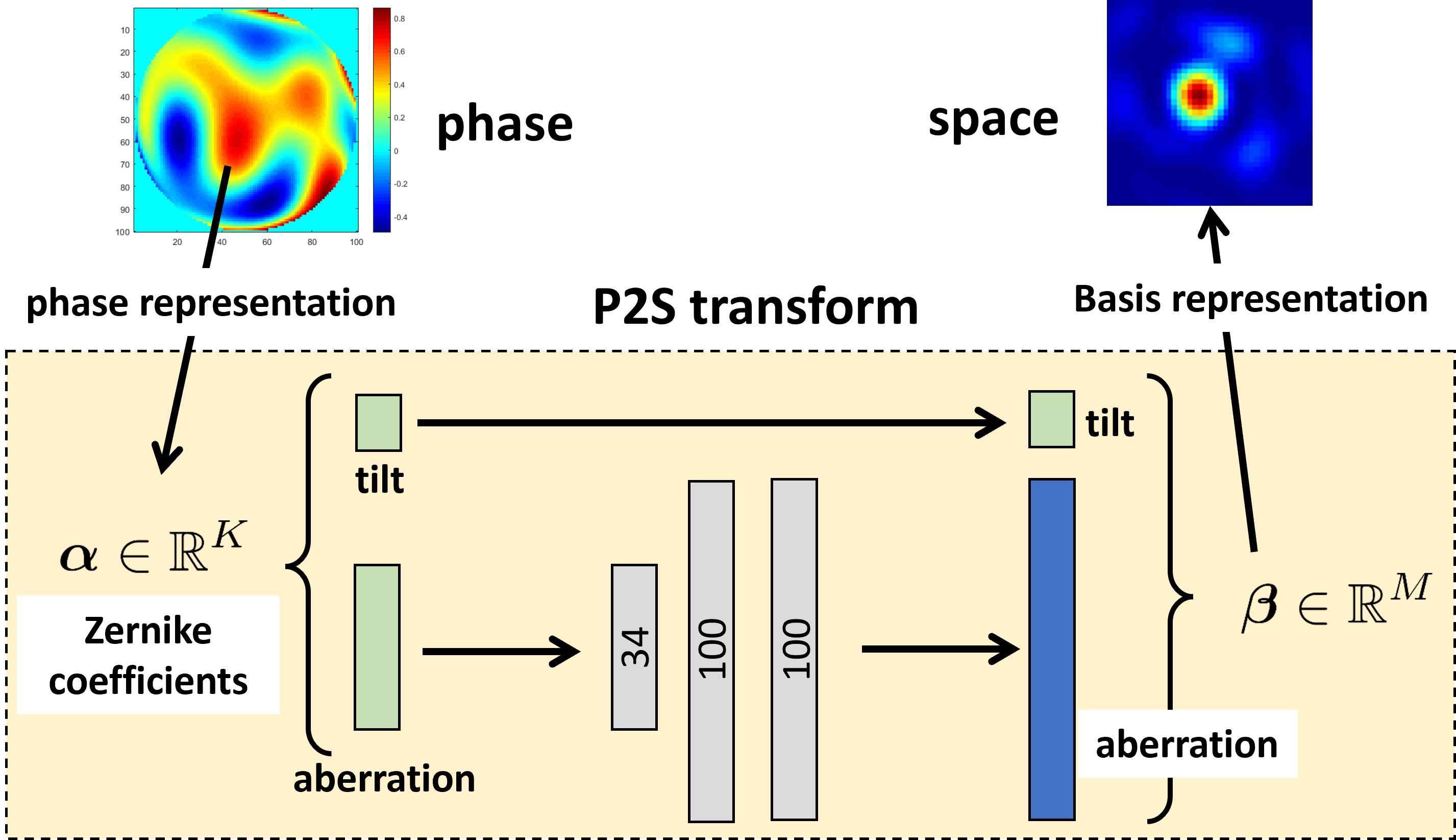}    \caption{Illustration of the Phase-to-Space transform. We bypass the computationally expensive PSF formation process by a learned mapping between the Zernike and spatial domain. We also note the sizes of the P2S layers here.}
    \label{fig:P2S}
\end{figure}

A schematic diagram of the P2S transform is shown in \fref{fig:P2S}. Given the two Zernike coefficients representing the tilts and the other Zernike coefficients representing the higher-order aberrations, the P2S transform uses the first two Zernike coefficients to displace the pixels, and uses the network to converts the last $K-2$ Zernike coefficients to $M$ basis representations.

The architecture of the P2S transform network consists of three fully connected layers as summarized in \fref{fig:P2S}. In terms of training, we re-use the 50,000 PSFs generated for Idea 2 train the P2S network. The training loss is defined as the $\ell_2$ distance between the predicted basis coefficients and the true coefficients (found offline by projecting the PSF onto the learned basis functions). Note that this network is light-weight because the P2S transform is performed \emph{per pixel}. For an image with a large field-of-view, the P2S network can be executed in parallel. Therefore, even with a $512\times512$ image, the entire transformation is done in a single pass.

\subsection{Interpolation across the grid}
We now address the computational difficulty for generating a dense set of Zernike coefficients $\valpha \in \R^K$ for a high resolution image. To accomplish this goal, we partition the image into a user-defined grid of anchor points, for example, a $64 \times 64$ grid. This grid corresponds to a correlation matrix of size $64^2 \times 64^2 = 4096 \times 4096$ which can be pre-computed. Following \fref{fig:basis}, 4096 sets of Zernike coefficients are drawn from the correlation matrix. To go from the grid of $64 \times 64$ anchor points to the full image, we interpolate the Zernike coefficients using bilinear interpolation. 

For generation of the anchor points, we implement the angle-of-arrival statistics according to \cite{Takato1995}, in conjunction with \cite{Basu_2015, chanan92}. The process is mathematically tedious but conceptually simple: One just needs to rewrite the entries of the correlation matrix in \cite{Chimitt2020} with the formula provided by \cite{Takato1995}. The output of the new correlation matrix is a set of spatially correlated Zernike coefficients.

\begin{figure}[ht]
    \centering
    \includegraphics[width=\linewidth]{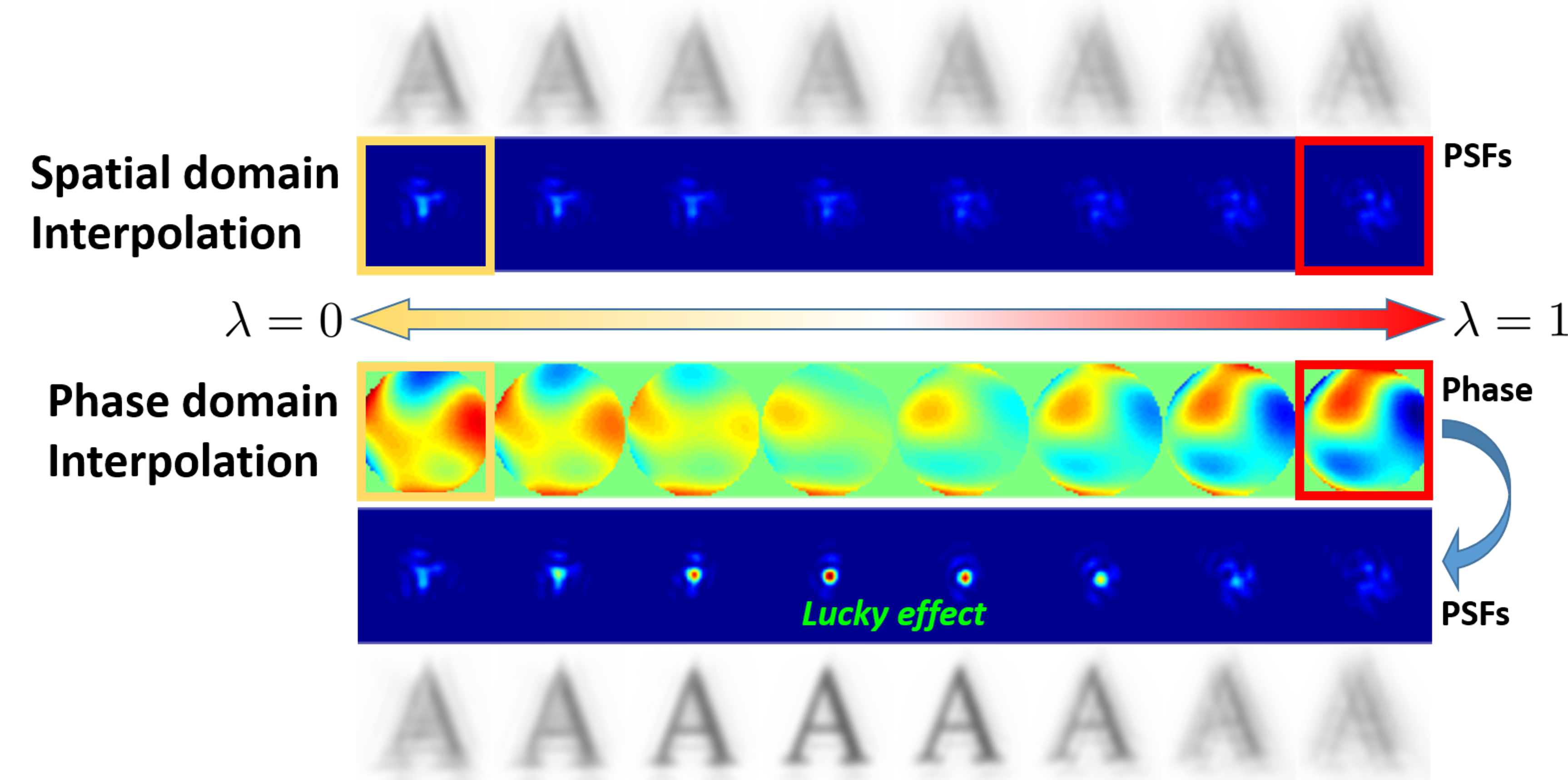}
    \caption{Comparison between the spatial interpolation scheme from \cite{Chimitt2020} and our interpolation in the phase domain. For both cases, we show the PSFs and example resultant images. [Top] Spatial interpolation of two PSFs is performed via $\lambda\vh_1 + (1-\lambda)\vh_2$, which is a superposition of the two PSFs. [Bottom] Phase interpolation is performed via $\lambda \phi_1 + (1-\lambda)\phi_2$. In this example, the superposition of the two phase functions will lead to a PSF with very mild phase distortion known as a lucky observation \cite{Fried78}. This lucky observation is absent in the spatial domain interpolation.
    }
    \label{fig: psf_interp}
\end{figure}

It is important to emphasize the difference between the way we interpolate and the interpolation used in \cite{Chimitt2020}. In \cite{Chimitt2020}, the interpolation is performed in the spatial domain where two PSFs are superimposed to generate a new PSF. In our simulator, we interpolate the Zernike coefficients to superimpose two phase functions. If the phase $\phi$ and the PSF $\vh$ is related by the P2S transform, $\phi \overset{\text{P2S}}{\longleftrightarrow} \vh$, it is important to note that for any $0 \le \lambda \le 1$,
\begin{align*}
\lambda \phi_1 + (1-\lambda)\phi_2 \;\; \cancel{\overset{\text{P2S}}{\longleftrightarrow}}\;\;
\lambda \vh_1 + (1-\lambda)\vh_2.
\end{align*}
Therefore, the interpolation used in \cite{Chimitt2020} is less justifiable. In \fref{fig: psf_interp} we illustrate the two interpolation schemes. We have selected a realistic and easily-observable case for illustration in which interpolation in the Zernike spaces generates a near-diffraction-limited PSF (the lucky effect \cite{Fried78}) but in the spatial domain is missed.

\subsection{Extension to color images}
Most deep neural networks today are designed to handle color images. To ensure that our simulator is compatible with these networks, we extend it to handle color. 

In principle, the spectral response of the turbulent medium is wavelength dependent, and the distortion must be simulated for a dense set of wavelengths. However, if the turbulence level is moderate, wavelength-dependent behavior of the Fried parameter is less significant for the visible spectrum (roughly 400nm to 700nm) when compared to other factors of the turbulence. 

\begin{figure}[ht]
	\centering
	\begin{tabular}{c c c}
\multicolumn{3}{c}{\includegraphics[width=0.9\linewidth]{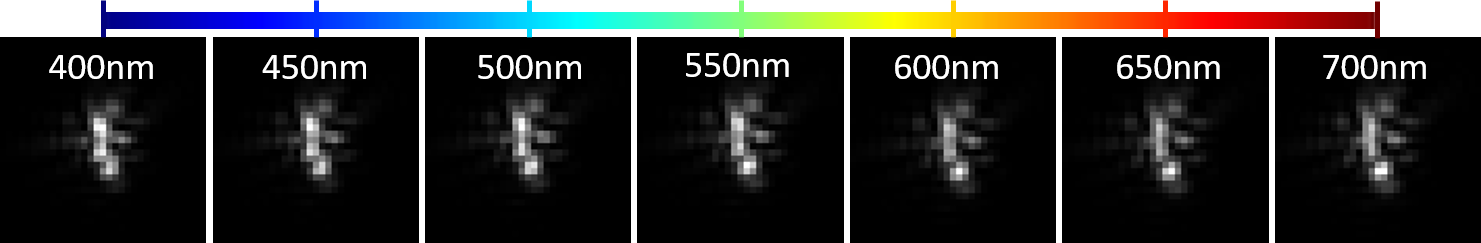}}\\
\multicolumn{3}{c}{(a)}\\
\includegraphics[width=0.3\linewidth]{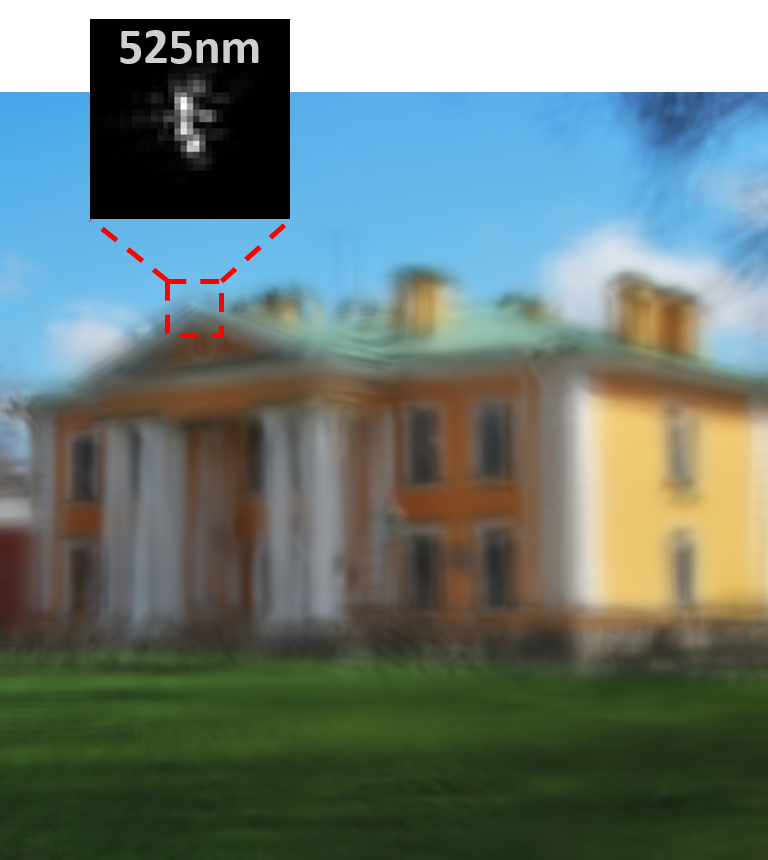}&
\hspace{-2ex}\includegraphics[width=0.3\linewidth]{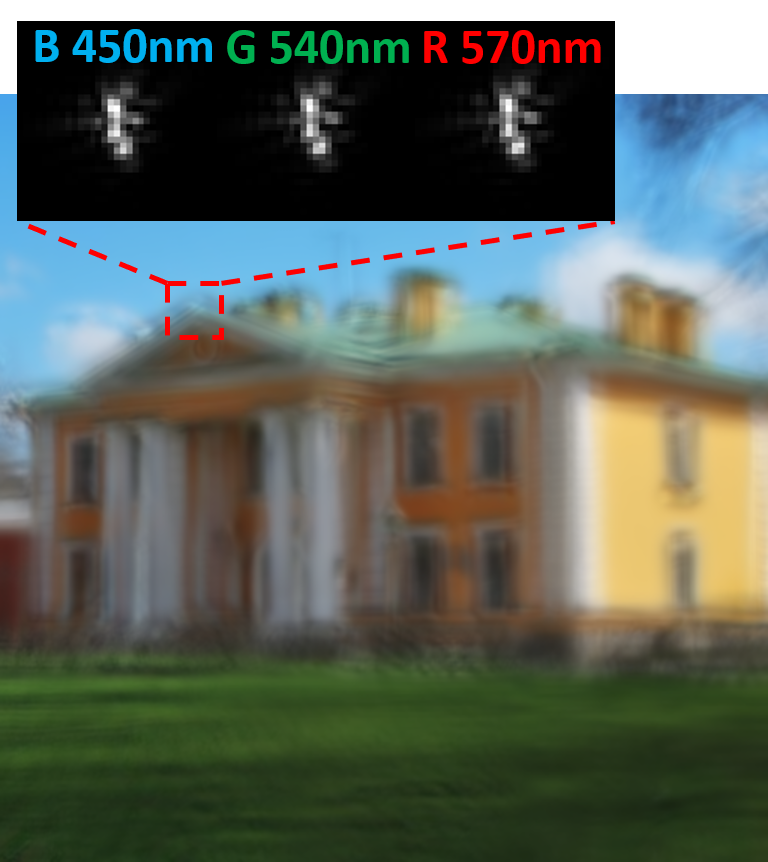}&
\hspace{-2ex}\includegraphics[width=0.3\linewidth]{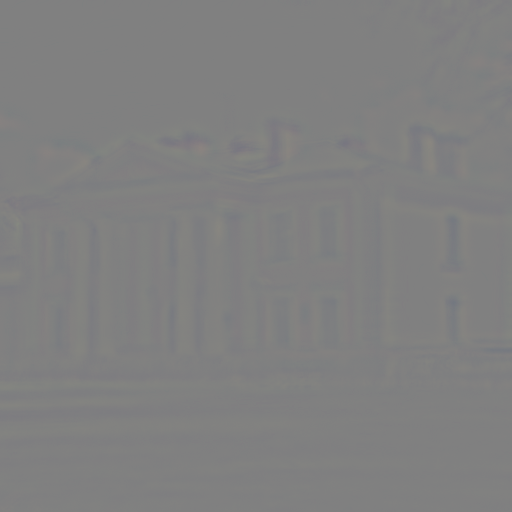}\\
(b)  & \hspace{-2ex}(c)  & \hspace{-2ex}(d) 
	\end{tabular}
	\caption{(a) PSFs across the visible spectrum. (b) Same distortion applied to three channels using center wavelength of the visible spectrum . (c) Wavelength dependent distortions applied to three channels. (d) Error map between (b) and (d). }
	\label{fig:color}
\end{figure}

To illustrate this observation, we show in \fref{fig:color} the individual PSFs for several wavelength from 400nm (blue) to 700nm (red). It is evident that the shape of the PSFs barely changes from one wavelength to another. In the same figure, we simulate two color images. The first image is simulated by using a single PSF (525nm) for the color channels (and displayed as an RGB image). The second image is simulated by considering 3 PSFs with wavelengths 450nm, 540nm, and 570nm. We note that (c) is a more realistic simulation but requires $3\times$ computation. However, the similar PSFs across the color makes difference is visually indistinguishable, as seen in (d). The small gap demonstrated in \fref{fig:color} suggests that we can simulate the RGB channels identically in such conditions.

\section{Experimental Evaluation}
Our experimental results consist of four parts: (i) Quantitative evaluation based on known turbulence statistics, (ii) visual comparison with real turbulence data, (iii) impact to deep neural network image reconstruction methods, (iv) run time comparison. Additionally, videos are included in the supplementary materials.

\subsection{Quantitative evaluation} 
\textbf{Evaluation schemes}. In the turbulence simulation literature, there are two standard ways to quantitatively evaluate a simulator: (i) the Z-tilt and the differential tilt statistics, and (ii) the short and long exposure statistics. For a simulator to be valid, it is necessary to match the simulated data with the theoretical curves. 

\textbf{Turbulence conditions}. To conduct this evaluation, we follow a similar setting as \cite{Chimitt2020} and \cite{Hardie2017}. The parameters of the turbulence are listed in the supplementary material.  

\begin{figure}[h]
	\centering
	\begin{tabular}{c}
		\includegraphics[width=0.85\linewidth]{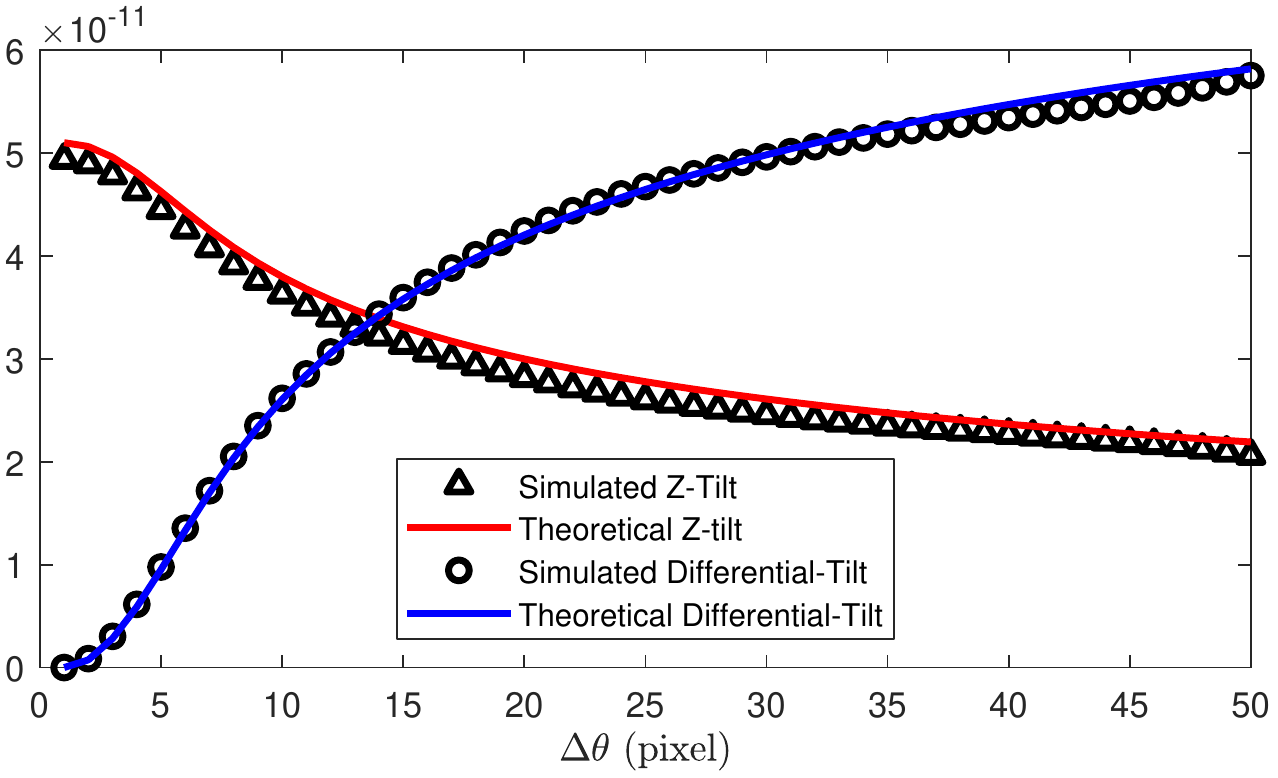}\\
	\end{tabular}
	\caption{The Z-tilt and differential-tilt statistics produced by our simulator match with the theoretical values.}
	\label{fig:tilt_stats}
\end{figure}

\textbf{Evaluation 1: Tilt statistics}. We first report the Z-tilt and the differential-tilt statistics. The Z-tilt and the differential-tilt statistics measure tilt correlation across the angle-of-arrivals. For example, the Z-tilt should drop as the angle-of-arrival increases, because two pixels that are far apart should have less (but non-zero) correlation. The results of the Z-tilt and the differential-tilt are shown in \fref{fig:tilt_stats}. It is evident that the tilt statistics of the proposed simulator matches well with the theoretical predictions.

\begin{figure}[h!]
	\centering
	\begin{tabular}{cc}
		\hspace{-2ex}\includegraphics[width=0.48\linewidth]{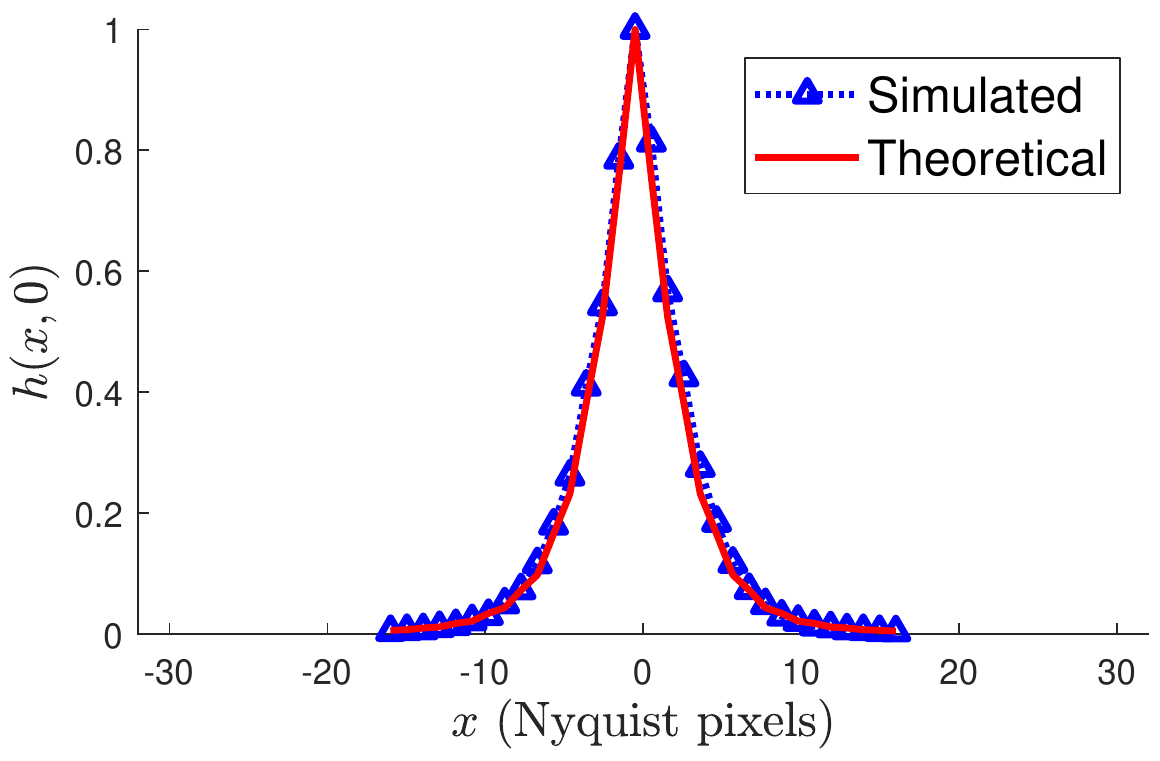}&
		\hspace{-2ex}\includegraphics[width=0.48\linewidth]{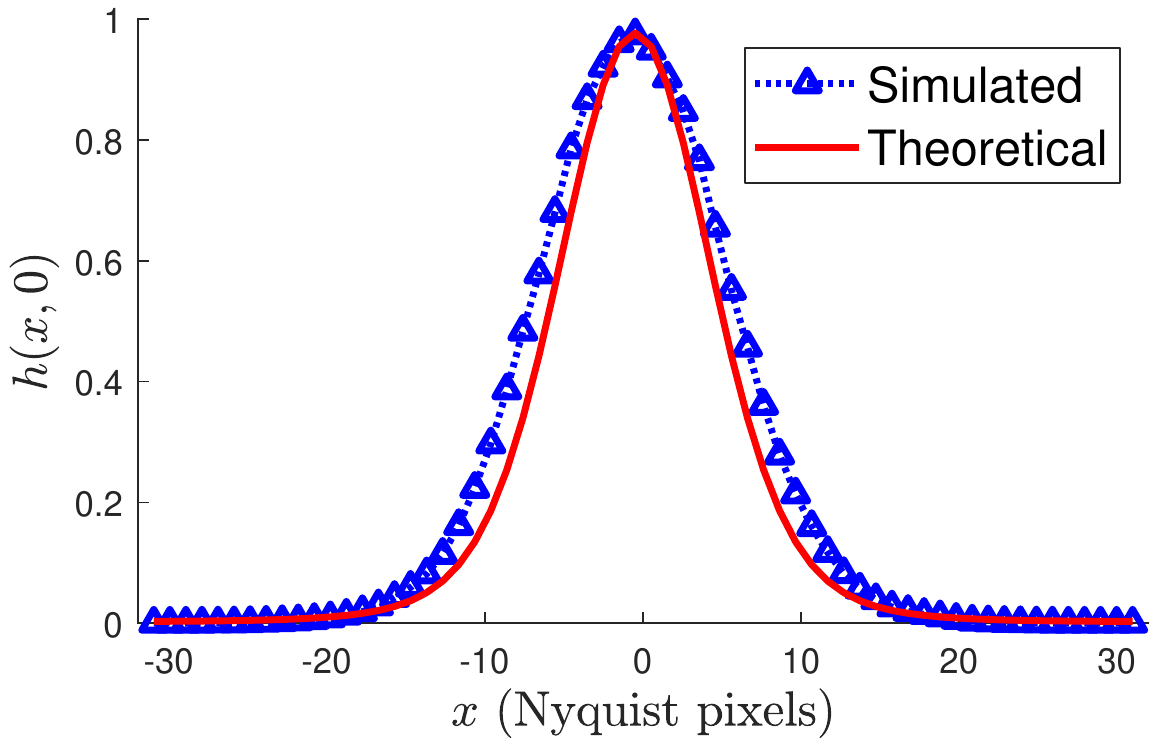}\\
		(a) Short exposure & (b) Long exposure
	\end{tabular}
	\caption{The short and long exposure PSFs produced by our simulator match with the theoretical PSFs.}
	\label{fig:Short_Long_Exp}
\end{figure}

\textbf{Evaluation 2: Long and short exposure}. We also analyze the long and short exposure (LE and SE, respectively) behavior of the generated PSFs. The LE PSF is a standard temporal average over the PSF realizations, while the SE is a temporal average over the centered PSFs. Since the LE includes pixel shifts, the spread of the LE PSF is larger than its SE counterpart. Furthermore, the SE is a valuable metric as it quantifies the blur the system experiences regardless of its shift behavior. We present these results in \fref{fig:Short_Long_Exp}, where we again see a match between the simulated and theoretical behavior.

\subsection{Visual comparison with real data} 
We emphasize the results of the previous quantitative discussion are significant statistically. However, visual comparisons with real data, while subjective, are an important consideration and serve as a useful reality check. In the following discussion, we present data simulated at the same optical parameters as those provided and show their real counterparts for visual comparison.

\begin{figure}[ht]
	\centering
	\begin{tabular}{c c c}\
\hspace{-2ex}\includegraphics[width=0.3\linewidth]{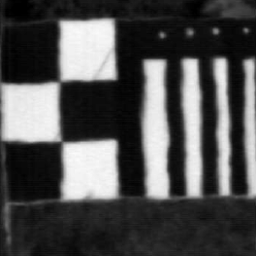}&
\hspace{-2ex}\includegraphics[width=0.3\linewidth]{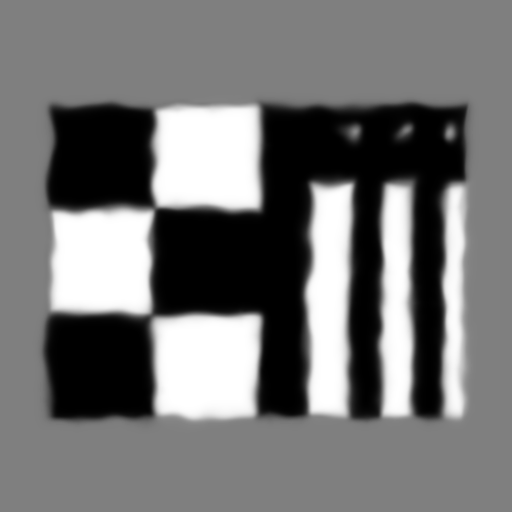}&
\hspace{-2ex}\includegraphics[width=0.3\linewidth]{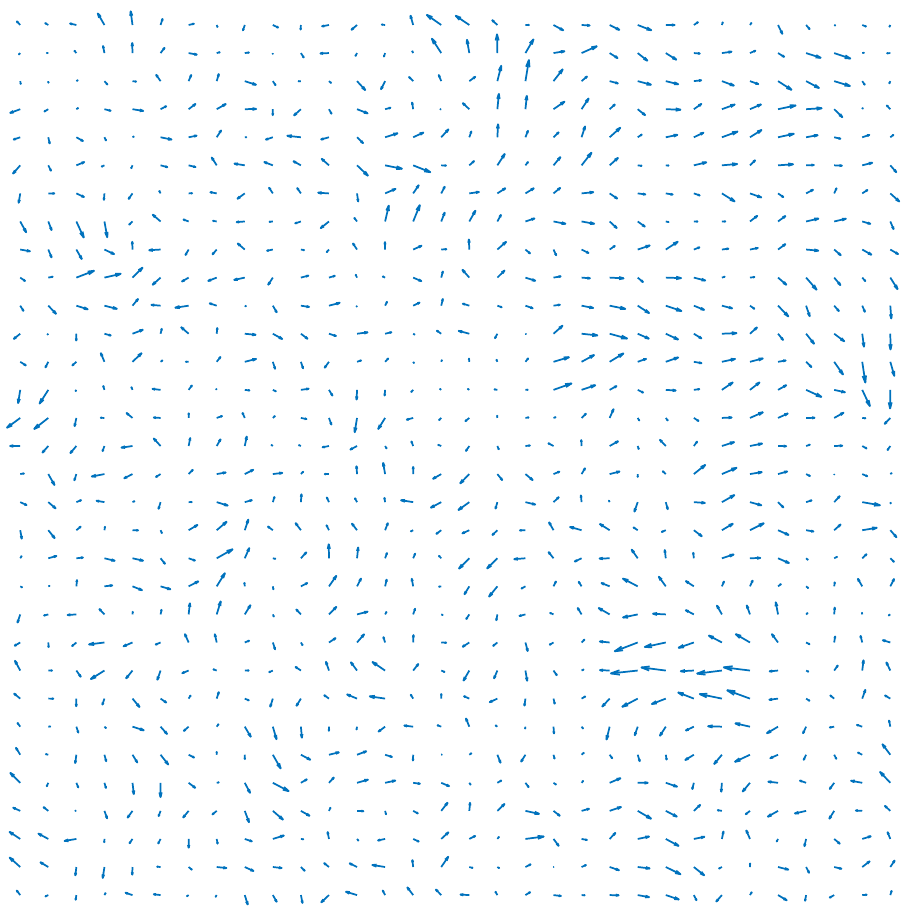}\\
\hspace{-2ex}\includegraphics[width=0.3\linewidth]{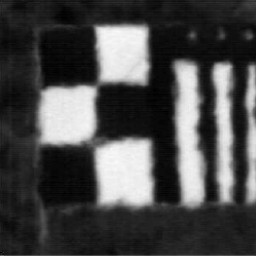}&
\hspace{-2ex}\includegraphics[width=0.3\linewidth]{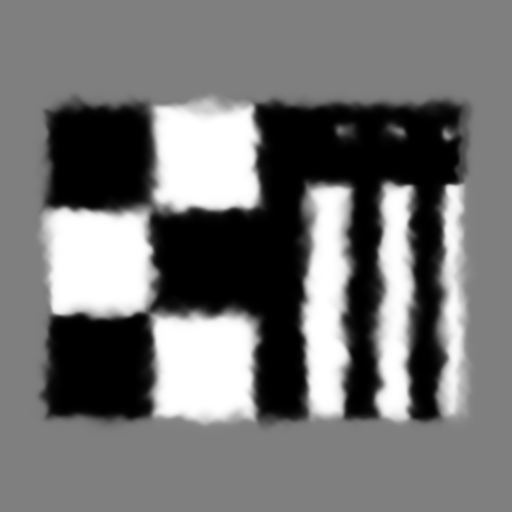}&
\hspace{-2ex}\includegraphics[width=0.3\linewidth]{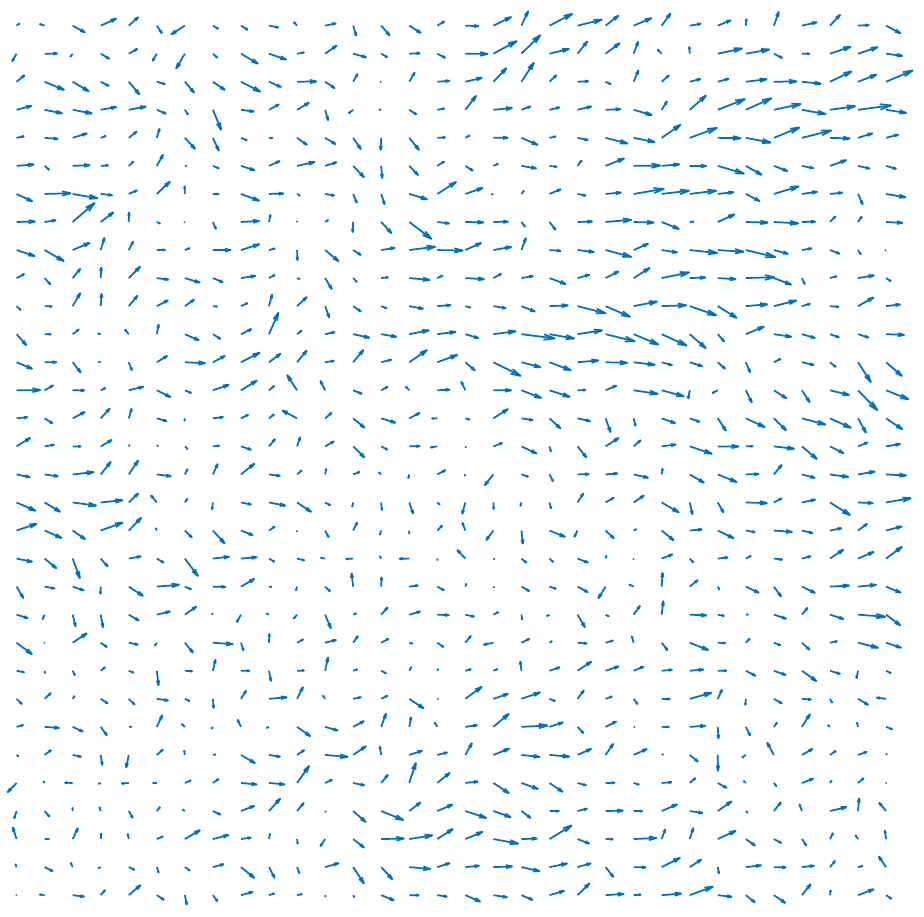}\\
\hspace{-2ex}\includegraphics[width=0.3\linewidth]{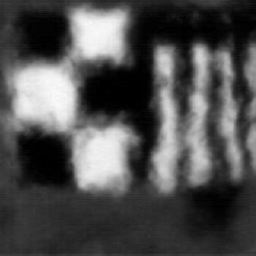}&
\hspace{-2ex}\includegraphics[width=0.3\linewidth]{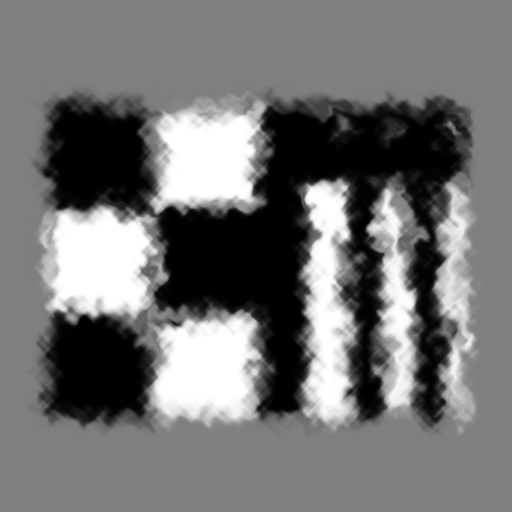}&
\hspace{-2ex}\includegraphics[width=0.3\linewidth]{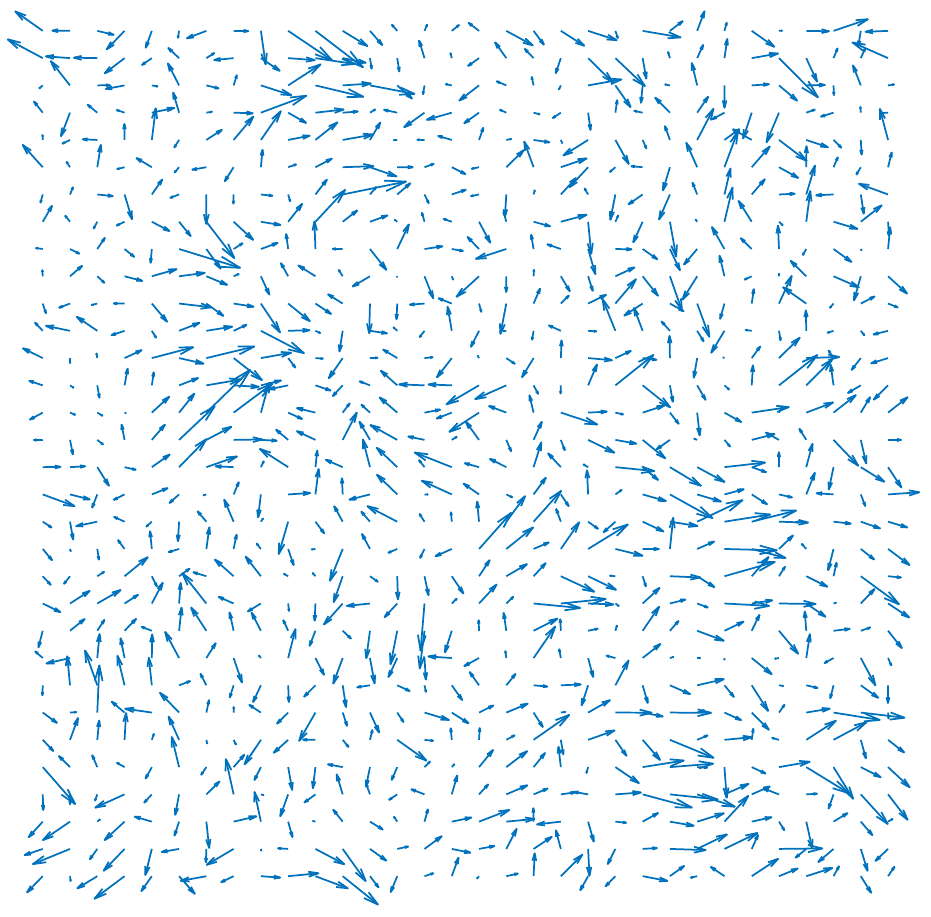}\\
(a) real & (b) simulated & (c) tilt map
	\end{tabular}
	\caption{Contrast balanced NATO RTG-40 dataset reported by \cite{nato,nato2}. The optical parameters are listed in supplementary materials. }
	\label{fig:field_data}
\end{figure}

\textbf{NATO dataset}. 
The carefully recorded NATO RTG40 dataset \cite{nato,nato2} contains both optical and estimated turbulence parameters. For these particular sets of images, the target is 1 km from the imaging system using passive visible light for imaging. Turbulence parameters were measured that help to evaluate what the appropriate turbulence level was at the time of taking the images. We select these parameters to use in our simulation technique, with comparisons shown in \fref{fig:field_data}.

In comparing simulated against their real counterparts, we can see a match in blur and shifting effects. At higher turbulence levels, there are some small observable differences, though we argue this is inherent to modeling just the phase in this type of problem as well as differences in illumination (e.g. digital representation of target pattern vs. illumination by the sun).

\begin{figure}[ht]
	\centering
	\begin{tabular}{ccc}
	    {(a) ground truth}&(b) real frame&(c) sim. frame \\
		\includegraphics[width=0.3\linewidth]{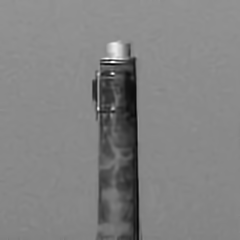}& \includegraphics[width=0.3\linewidth]{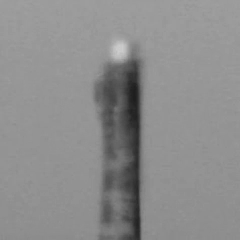} & \includegraphics[width=0.3\linewidth]{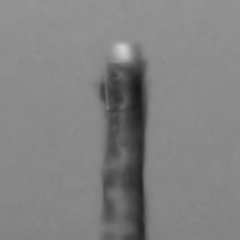} \\
		\includegraphics[width=0.3\linewidth]{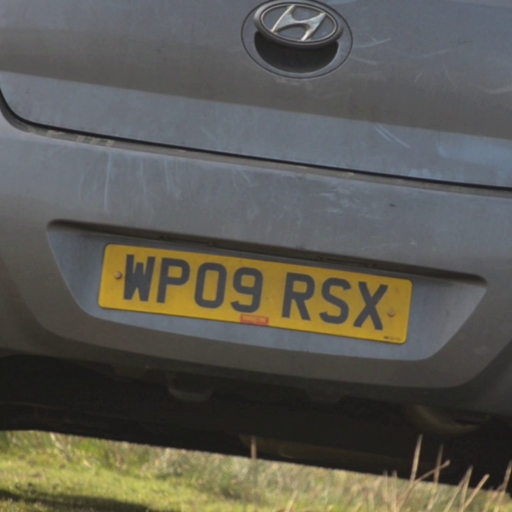} & \includegraphics[width=0.3\linewidth]{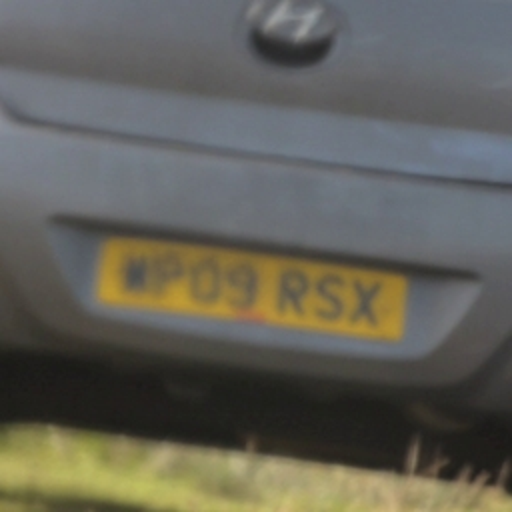} & \includegraphics[width=0.3\linewidth]{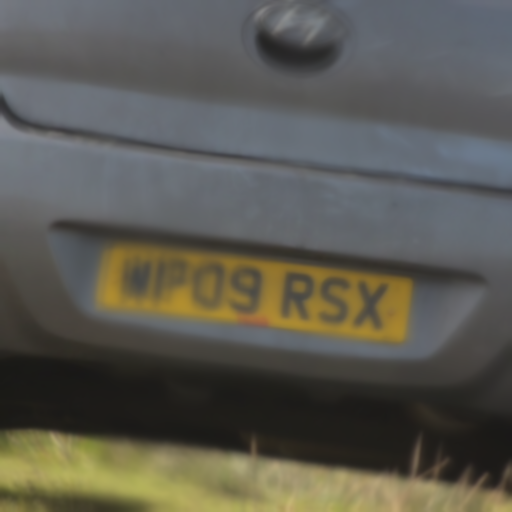}
	\end{tabular}
	\caption{Visual comparison of simulated and real turbulence data. With comparing individual frames, we can see similar blurring and warping effects. }
	\label{fig:visual}
\end{figure}

\textbf{Datasets used in \cite{Hirsch2010} and \cite{Anantrasirichai2013}}. 
In addition to the NATO dataset, there are also those used in \cite{Hirsch2010} and \cite{Anantrasirichai2013}. The images in \fref{fig:visual} show a method of collecting turbulence data that uses stream of gas in front of the camera to produce images at different turbulence levels. While this is a different scenario than the typical long-distance imaging sequences, this data serves as a decent proxy and is useful as it is easier to collect and can provide ground truth by simply turning the gas system off. We present for visual comparison the results in \fref{fig:visual} and note the similarity in random draws vs. observations.

\begin{figure*}[h!]
	\centering
	\begin{tabular}{c c c c c}
	
		\includegraphics[width=0.18\linewidth]{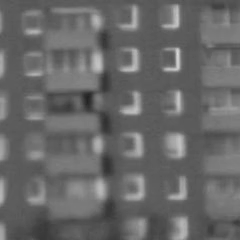}&
		\hspace{-2ex}\includegraphics[width=0.18\linewidth]{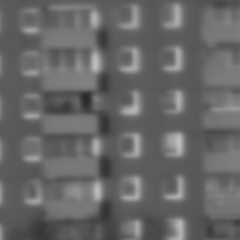}& \hspace{-2ex}\includegraphics[width=0.18\linewidth]{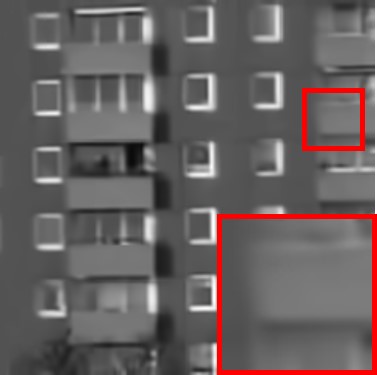}&
		\hspace{-2ex}\includegraphics[width=0.18\linewidth]{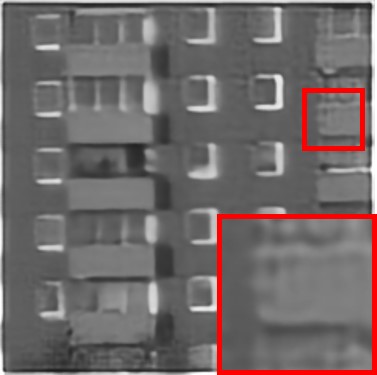}&
		\hspace{-2ex}\includegraphics[width=0.18\linewidth]{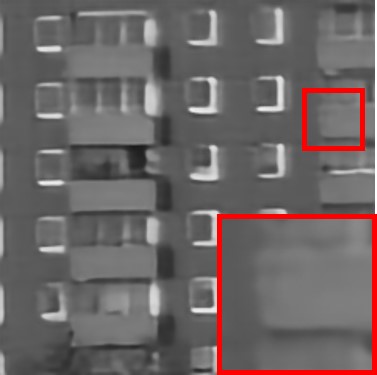}\\

		\includegraphics[width=0.18\linewidth]{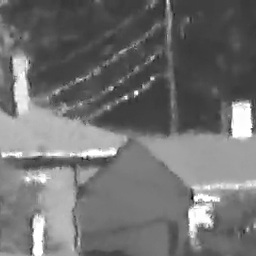}&
		\hspace{-2ex}\includegraphics[width=0.18\linewidth]{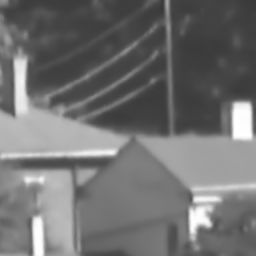}& \hspace{-2ex}\includegraphics[width=0.18\linewidth]{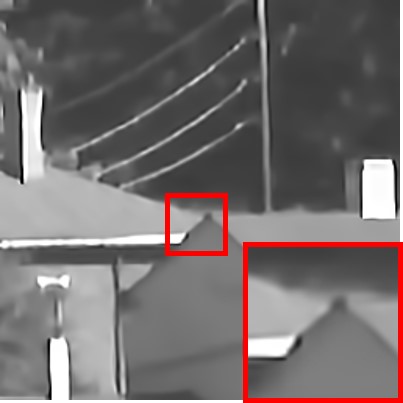}&
		\hspace{-2ex}\includegraphics[width=0.18\linewidth]{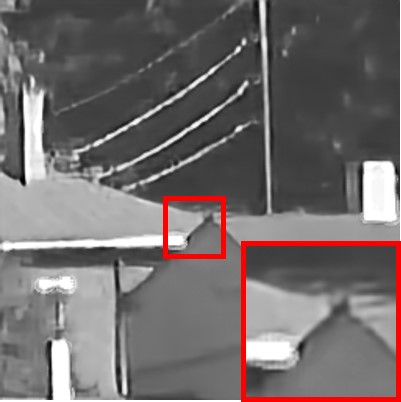}&
		\hspace{-2ex}\includegraphics[width=0.18\linewidth]{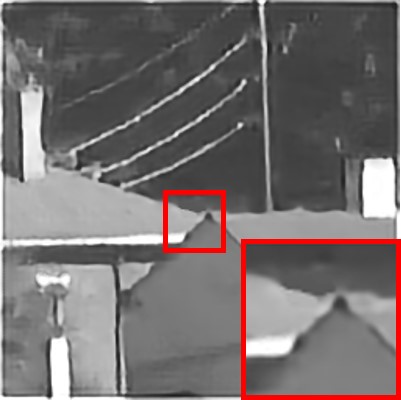}\\
		
		(a) Input (real) & \hspace{-2ex} (b) Temp. Avg. & \hspace{-2ex} (c) Mao et al. \cite{mao_tci} & \hspace{-2ex} (d) \cite{Lau2021_sim}+U-Net & \hspace{-2ex} (e) Ours+U-Net\\
	\end{tabular}
	\caption{Image reconstruction using \textbf{real} data (so ground truth is not available). For (d) and (e), we train a UNet using data synthesized by \cite{Lau2021_sim} and our simulator, respectively. Notice the artifacts in (d).}
	\vspace{-2ex}
	\label{fig:recon}
\end{figure*}

\subsection{Impact on training deep networks}

We conduct an experiment to demonstrate the impact of the proposed simulator on a multi-frame turbulence image reconstruction task. The goal of this experiment is to show that a deep neural network trained with the data synthesized by the proposed simulator outperforms the same network trained with the data generated by simulators that are less physically justified. 

To demonstrate the impact of the simulator, we do not use any sophisticated network structure or training strategy. Our network has a simple U-Net architecture \cite{Unet} with 50 input channels and is trained with an MSE loss for 200 epochs. The network is trained with 5000 simulated sequences, where each sequence contains 50 degraded frames. The ground truth images used for simulation are obtained from the Places dataset \cite{placesdataset}. The sequences are simulated with a turbulence level $D/r_0$ uniformly sampled from [1,8].  

For comparison, we train the same network using a simulation technique proposed by Lau et al. \cite{Lau2021_sim}. This simulator has been used in several recent works \cite{Lau2021,Yasarla2020}. To ensure a fair comparison, we perform a uniform sweep for the Gaussian blur ($\sigma^2$ sampled from [1, 3]) and tilt strength (sampled from [0.1, 0.4]). As a reference, we also report the results of a deterministic (non-learning based) state-of-the-art reconstruction method by Mao et al. \cite{mao_tci}. 

Two qualitative reconstruction results are shown in \fref{fig:recon}. It can be seen that the network trained with proposed simulator has performance close to state-of-the-art. Visible artifacts are generated from the network trained with \cite{Lau2021_sim}. We also include a quantitative evaluation, where a split-step simulator \cite{Hardie2017} is used to generate 30 testing sequences under low, medium, and high ($D/r_0 = 1.5\text{, } 3\text{,\ and } 4.5$). PSNR values are reported in Table \ref{tab:recon_psnr}. It is worth nothing that the network trained with the data synthesized by our simulator achieves a comparable performance to the state-of-the-art. 

\begin{table}[]
    \centering
    \begin{tabular}{c c c c}
        \hline\hline
         $D/r_0$ & Mao et al. \cite{mao_tci}& Ours+U-Net & \cite{Lau2021_sim}+U-Net\\
        \hline
        1.5 & 27.33dB & 27.18dB & 26.59dB \\
        3.0 & 27.04dB & 26.98dB & 26.11dB \\
        4.5 & 25.85dB & 26.01dB & 25.40dB \\
        \hline
    \end{tabular}
    \vspace{1ex}
    \caption{PSNR values of the reconstruction results, averaged over 30 testing sequences. The testing data is \textbf{synthesized} by the split-step propagation method \cite{Hardie2017}. }
    \vspace{-2ex}
    \label{tab:recon_psnr}
\end{table}

\subsection{Run time}
Finally, we compare the run time of the proposed method with several existing methods \cite{ Chimitt2020,Hardie2017, Lau2021_sim}. The simulators are run on a computing cluster node with Intel Xeon ``Sky Lake" processors (16 cores) and a Tesla V100 GPU. We use $16\times16$ PSF grid for \cite{Chimitt2020}, which is comparable to our initial PSF grid. The for-loop in \cite{Lau2021_sim} is executed 1000 times as suggested by the authors. The run time of \cite{Hardie2017} is reported by the authors. The required time to process a $256\times256$ frame is reported in Table \ref{tab:runtime}. The proposed method offers $300\times$--$1000\times$ speed up compared to Hardie et al. \cite{Hardie2017}.

\begin{table}[]
    \centering
    \begin{tabular}{c c c c}
        \vspace{-2ex}\\
        \hline\hline
        Reference & Method &  CPU (s) &  GPU (s) \\
        \hline
        Hardie et al. \cite{Hardie2017} & split-step & 119.63  & 24.36  \\
        Chimitt-Chan \cite{Chimitt2020} & collapsed & 5.88  & N/A\\
        Lau et al. \cite{Lau2021_sim} & subsampling & 3.13 & N/A \\
        Ours & P2S & 0.35 & 0.026 \\
        \hline
    \end{tabular}
    \vspace{1ex}
    \caption{Average run time for each method to process a $256\times256$ frame. Unit are in seconds. }    
    \label{tab:runtime}
\end{table}

\section{Conclusion}
The simulation approach towards imaging through atmospheric turbulence we have presented in this work has desirable advantages over existing methods. The key innovation of the P2S transform network allows for significant speedup and additional reconstruction utility. With respect to deep-learning based reconstruction, the outlined approach allows for the generation of large amounts of training data not previously feasible. Additionally, the ability to use the simulation approach as a differentiable module in a neural-network suggests additional benefit towards reconstruction. Finally, we expect the ability to produce statistically accurate data far more efficiently will allow for further statistical analysis of turbulent imaging properties through numerical analysis methods not previously possible.

\subsection*{Acknowledgement}
The work is supported, in part, by the National Science Foundation under the grants CCF-1763896 and ECCS-2030570.

\clearpage
\newpage
{\small
\bibliographystyle{ieee_fullname}
\bibliography{egbib}
}

\end{document}